\documentclass[amsmath,twocolumn,prb,superscriptaddress,longbibliography]{revtex4-1}

\usepackage[colorlinks=true,allcolors=blue]{hyperref}
\usepackage{amsmath}
\usepackage{amssymb}
\usepackage{bm}
\usepackage{xcolor}
\usepackage[utf8]{inputenc}
\usepackage{graphicx}
\usepackage{dsfont} 
\usepackage{subfigure}

\usepackage{amsmath}
\usepackage{amsfonts}
\usepackage{braket}
\usepackage{dsfont}

\newcommand{\sgn}[1]{\text{sgn}(#1)}
\newcommand{\bk}{\mathbf{k}}
\newcommand{\bq}{\mathbf{q}}
\newcommand{\br}{\mathbf{r}}

\newcommand{\re}{\mathrm{e}}
\newcommand{\rd}{\mathrm{d}}

\newcommand{\dg}{\dagger}

\begin{document}

\title{Coupled wire construction of a topological phase with chiral tricritical Ising edge modes}
\author{Chengshu Li}
\email{chengshu@phas.ubc.ca}
\affiliation{Department of Physics and Astronomy \& Stewart Blusson Quantum Matter Institute, University of British Columbia, Vancouver, British Columbia  V6T 1Z1, Canada}
\author{Hiromi Ebisu}
\affiliation{Department of Condensed Matter Physics, Weizmann Institute of Science, Rehovot 76100, Israel}
\author{Sharmistha Sahoo}
\affiliation{Department of Physics and Astronomy \& Stewart Blusson Quantum Matter Institute, University of British Columbia, Vancouver, British Columbia  V6T 1Z1, Canada}
\author{Yuval Oreg}
\affiliation{Department of Condensed Matter Physics, Weizmann Institute of Science, Rehovot 76100, Israel}
\author{Marcel Franz}
\email{franz@phas.ubc.ca}
\affiliation{Department of Physics and Astronomy \& Stewart Blusson Quantum Matter Institute, University of British Columbia, Vancouver, British Columbia  V6T 1Z1, Canada}

\begin{abstract}
Tricritical Ising (TCI) phase transition is known to occur in several interacting spin and Majorana fermion models and is described in terms of a supersymmetric conformal field theory (CFT) with central charge $c=7/10$. The field content of this CFT is highly nontrivial and includes among its primary fields the Fibonacci anyon, making it of potential interest to strategies seeking to implement fault-tolerant topological quantum computation with non-Abelian phases of matter. In this paper we explore the possibility that a TCI CFT can occur at the edge of a gapped two-dimensional topological state as a stable phase. We discuss a possible realization of  this 2D phase based on a coupled-wire construction  using the Grover-Sheng-Vishwanath chain model 
of Majorana zero modes coupled to Ising spins which is known to  undergo the TCI phase transition. From the combined analysis using mean-field theory, conformal field theory and density matrix renormalization group (DMRG) on 2- and 4-leg ladders, we find that the left- and right-moving gapless TCI modes become spatially separated and reside on two opposite edges of the system, forming a precursor of the required 2D topological phase. 
\end{abstract}

\date{\today}

\maketitle

\section{Introduction}

Since the advent of the family of quantum Hall effects [\onlinecite{Klitzing1980,Tsui1982,Laughlin1983}], edge physics in two-dimension (2D) has attracted increasing interest from both the experimental and theoretical condensed matter communities. Experimentally, edge modes result in distinctive behavior in transport [\onlinecite{Klitzing1980}, \onlinecite{Tsui1982}] and tunneling [\onlinecite{Chang1996}] measurements, leading to extraordinarily precise metrology of fundamental constants of physics [\onlinecite{Codata}]. Theoretically, this can be viewed as manifestation of the deep relation between the edge and the bulk, known as the bulk-boundary correspondence [\onlinecite{Hasan2010}, \onlinecite{Qi2011}], which allows one to infer bulk properties from edge measurements and vice versa.

One powerful theoretical method for probing the edge properties is the coupled wire construction [\onlinecite{Kane2002,Teo2014,Klinovaja2014,Stoudenmire2015,Kane2017,Sagi2017,Laubscher2019,Yang2020}], where a 2D system is built from  an array of one-dimension (1D) chains. Thanks to a suite of  powerful tools ranging from bosonization to conformal field theory, 1D models are better understood and sometimes allow exact solutions unavailable in their  higher dimensional counterparts. Studying an anisotropic assembly of coupled 1D chains  then brings insights from the constituent 1D models into the 2D model of interest. In particular, the chiral 1D edges of the 2D system are directly related to the 1D building blocks.

In this work we apply these ideas to 1D systems that exhibit the TCI phase transition point in their phase diagram. Our goal is to understand if the chiral edge modes of the stable 2D gapped phase which we attempt to construct here are described by e.g. the left moving holomorphic sector of the TCI critical point. Because the primary fields of TCI CFT contain a Fibonacci anyon, the bulk-boundary correspondence guarantees that Fibonacci anyon will also exist as a gapped excitation of the bulk system [\onlinecite{Hu2018}]. There are several concrete microscopic models of 1D spins and interacting Majorana fermions that exhibit the TCI point [\onlinecite{Friedan1985, Zamolodchikov1986, Kastor1989, Mussardo}]. These include the original spin-1 model [\onlinecite{Blume1966,Capel1966,Blume1971}], interacting Majorana fermion models in 1D chain [\onlinecite{Rahmani2015a,Rahmani2015b,O'Brien2018,Sannomiya2019}] and ladder [\onlinecite{Zhu2016}], and the Grover-Sheng-Vishwanath (GSV) model involving Majorana fermions and spins [\onlinecite{Grover2014}]. 
In the following we focus on the GSV model because the TCI point can be reached by tuning a single model parameter and the transition occurs at an intermediate coupling strength making DMRG simulations relatively well behaved.

\begin{figure}
    \centering
    \includegraphics[width=0.5\textwidth]{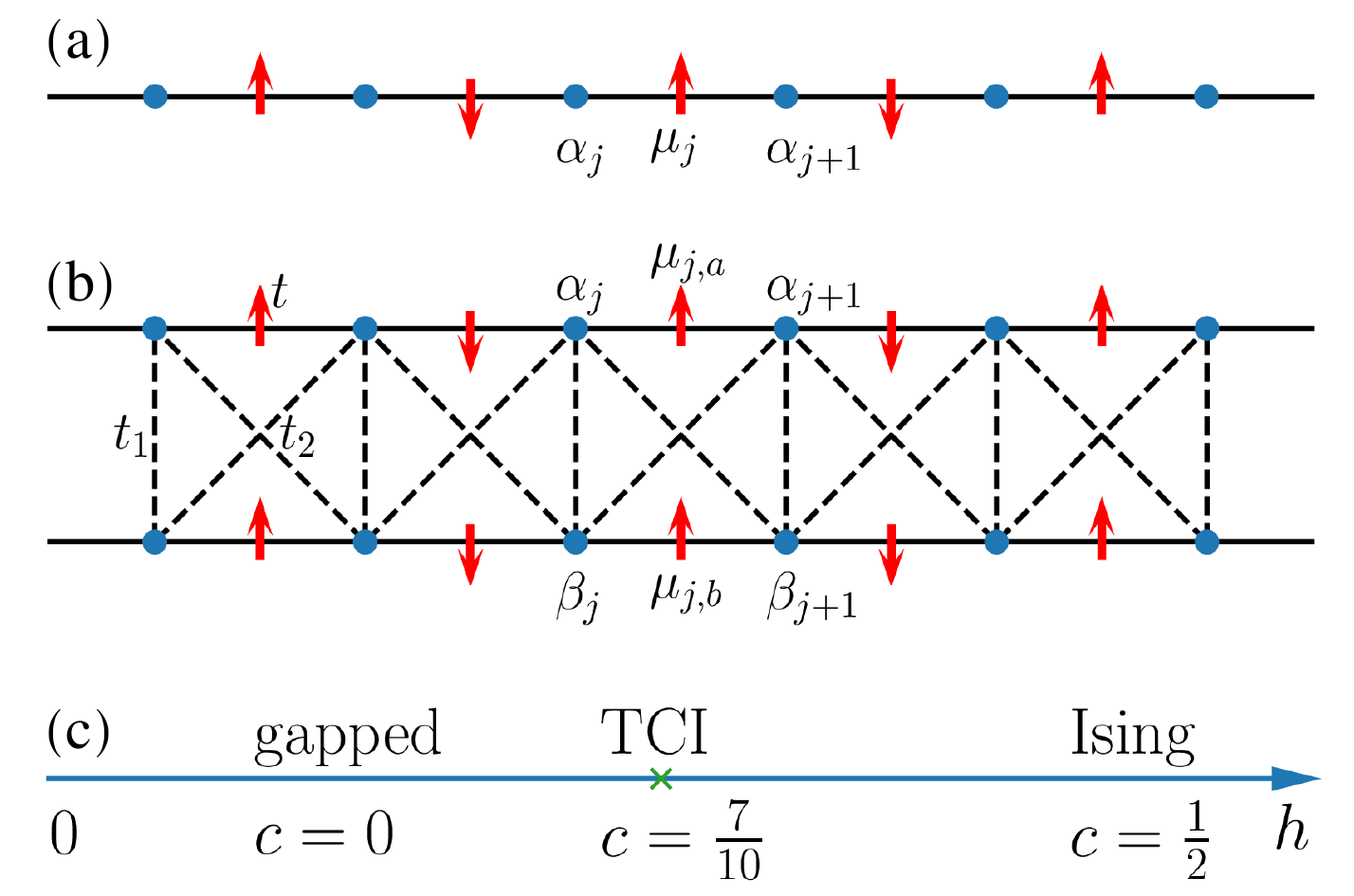}
    \caption{(a) Lattice structure of the GSV model, Eq.~(\ref{eq:H_GSV}). (b) Lattice structure of the ladder model, Eq.~(\ref{eq:H_ladder}). (c) The phase diagram of both models.}
    \label{fig:1}
\end{figure}

The GSV model comprises Majorana fermions $\alpha_j$ on sites and spin-$\frac{1}{2}$ degrees of freedom $\mu_j$ living on bonds of  a 1D chain as depicted in Fig.~\ref{fig:1}(a). The Majorana hoppings are modulated by the spins which are in turn described by the transverse field Ising model. The Hamiltonian $ H_\mathrm{GSV}=H_\mathrm{M}+H_\mathrm{s}+H_{\mathrm{Ms}}$, of the model reads
\begin{equation}
\begin{split}
&H_\mathrm{M}=it\sum_j\alpha_j\alpha_{j+1},\\
&H_\mathrm{s}=J\sum_j\mu_{j}^z\mu_{j+1}^z-h\sum_j\mu_{j}^x,\\
&H_{\mathrm{Ms}}=-igt\sum_j\alpha_j\alpha_{j+1}\mu_{j}^z,
\end{split}\label{eq:H_GSV}
\end{equation}
where $t,h,g,J>0$. The system is gapped when $h\rightarrow0$ because the spin chain symmetry is spontaneously broken giving a mass to the Majorana chain. In the limit $h\rightarrow \infty$ the average of $\mu_j^z$ is zero leaving the Majorana chain gapless, which is dual to a transverse field Ising critical spin chain. The two stable phases are separated by the TCI phase transition [\onlinecite{Grover2014}] that occurs for $h=h_c(g)$, with central charge $c=\frac{7}{10}$, see Fig.~\ref{fig:1}(c).

The ladder model we consider is built from GSV chains where we allow Majorana fermion hopping between chains  as depicted in Fig.~\ref{fig:1}(b). The Hamiltonian of the 2-leg ladder $H_\mathrm{ladder}=H_\mathrm{M}+H_\mathrm{s}+H_{\mathrm{Ms}}, $ is given by
\begin{equation}
\begin{split}
&H_\mathrm{M}=it\sum_j(\alpha_j\alpha_{j+1}-\beta_j\beta_{j+1})+it_1\sum_j\beta_j\alpha_j\\
&\qquad+it_2\sum_j(\alpha_j\beta_{j+1}-\beta_j\alpha_{j+1}),\\
&H_\mathrm{s}=J\sum_j\mu_{j,a}^z\mu_{j+1,a}^z-h\sum_j\mu_{j,a}^x+(a\rightarrow b),\\
&H_{\mathrm{Ms}}=-igt\sum_j(\alpha_j\alpha_{j+1}\mu_{j,a}^z-\beta_j\beta_{j+1}\mu_{j,b}^z).
\end{split}\label{eq:H_ladder}
\end{equation}
Using a  combination of analytical and numerical methods, we find that the 2-leg ladder model goes through a TCI transition when we tune $h$ from 0 to $\infty$, just like the single chain. In addition we show that the left- and right-moving TCI modes spatially separate, such that they reside on the upper and the lower leg of the ladder, respectively. A similar behavior is seen in the 4-leg ladder although here our numerics are less reliable. Collectively these results suggest that ladders composed of weakly coupled GSV chains can be viewed as a precursor of a 2D topological phase with a fully gapped bulk and chiral gapless TCI modes bound to its edges. 

The paper is organized as follows. We first briefly discuss the relevant non-interacting models, both to serve as a mean-field treatment of the full model and to obtain intuition of the coupled wire construction. In particular, we examine the edge modes with analytical and numerical approach. Then we show numerical results of the full model which lend support to the analytical analysis. We conclude by commenting on generalizations to multi-leg and 2d models.

\section{Non-interacting models}

Much of our intuition of the coupled wire construction leans on the non-interacting case. Coincidentally it turns out that some of the interesting physics in the interacting model can be understood based on  a simple mean-field treatment. Therefore we provide a brief discussion of the relevant mean-field theory in this section. We start with a Majorana chain model 
\begin{equation}
H_\mathrm{chain,0}=i\sum_j(t+m(-1)^j)\alpha_j\alpha_{j+1}.\label{eq:H_chain0}
\end{equation}
This can be viewed as a mean-field version of the GSV model as follows. We assume that in Eq.~(\ref{eq:H_GSV}) the spin $\mu^z$ in $H_\mathrm{Ms}$ is not dynamical but enters as a mean-field parameter $m$ which is determined by the Ising Hamiltonian $H_\mathrm{s}$. As $h\rightarrow0$, the spins are antiferromagnetically ordered, and $m\neq0$ leaving the fermion system gapped. On the other hand as $h\rightarrow\infty$ we have $m=0$ and the fermion system becomes critical. When the dynamics of the spin degrees of freedom is restored  we expect a TCI phase transition between the two phases with spins providing the gapless bosonic excitations required by the supersymmetric nature of the transition. This analysis matches well with DMRG calculations as reported in Ref.~\onlinecite{Grover2014}.

Now we couple two such chains into a ladder as shown in Fig.~\ref{fig:1}(b), with phases consistent with the Grosfeld-Stern rule [\onlinecite{Grosfeld2006}, \onlinecite{Liu2015}]. The Hamiltonian reads
\begin{equation}
\begin{split}
&H_\mathrm{ladder,0}=i\sum_j(t+m(-1)^j)(\alpha_j\alpha_{j+1}-\beta_j\beta_{j+1})\\
&\qquad+it_1\sum_j\beta_j\alpha_j+it_2\sum_j(\alpha_j\beta_{j+1}-\beta_j\alpha_{j+1}).
\end{split}\label{H_ladder0}
\end{equation}
The spectrum is gapless when $t_2=\sqrt{t_1^2/4+m^2}$. If we fix $t_1=2t_2$, and again assume $m$ to be a mean-field for $\mu^z$, a similar argument as for  the chain model follows. In particular, we expect a TCI phase transition at a critical point $h_c$ in the full model Eq.~(\ref{eq:H_ladder}). We will have more to say about the interacting model in the next section.

\begin{figure}
    \centering
    \includegraphics[width=0.48\textwidth]{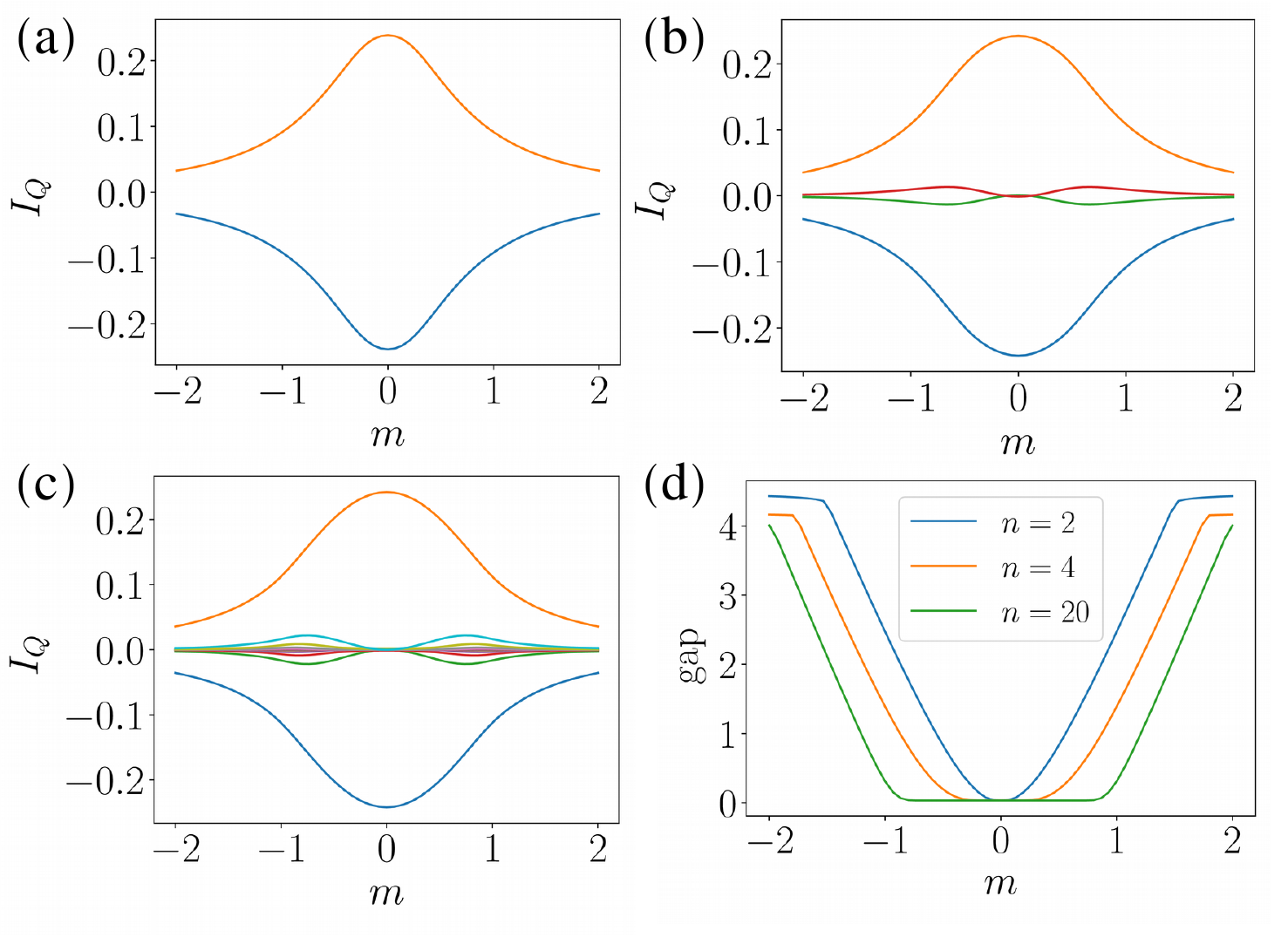}
    \caption{(a-c) The behavior of the thermal currents as a function of $m$ in the non-interacting model for (a) two-leg, (b) four-leg, and (c) 20-leg systems. In all the figures we set $t=t_1=2t_2=1$ and color the top (bottom) leg as blue (orange). (d) The energy gap of the non-interacting model as a function of $m$. The plateau at zero energy suggests the existence of gapless edge modes.}
    \label{fig:2}
\end{figure}

Another useful perspective comes from the continuum limit where we focus on the long-wavelength low-energy behavior. The corresponding Hamiltonian is derived by expanding the Majorana fields near the nodal point in the momentum and retaining only terms to linear order in the momentum.  
This amounts to setting
\begin{equation}
\alpha_{2j}\rightarrow \alpha_e,\ 
\alpha_{2j+1}\rightarrow \alpha_o,\ 
\beta_{2j}\rightarrow \beta_e,\ 
\beta_{2j+1}\rightarrow \beta_o,\label{5}
\end{equation}
and defining  the ``chiral'' basis
\begin{equation}
\alpha_{R/L}=\frac{1}{\sqrt{2}}(\alpha_o\pm\alpha_e),\ 
\beta_{R/L}=\frac{1}{\sqrt{2}}(\beta_o\mp\beta_e).\label{6}
\end{equation}
For simplicity we focus on the case $t_1=2t_2$, and the Hamiltonian becomes
\begin{equation}
\begin{split}
H&=i\int\rd x(t-m)(\alpha_R\partial\alpha_R+\beta_R\partial\beta_R-\alpha_L\partial\alpha_L\\
&\qquad-\beta_L\partial\beta_L)+2t_2(\alpha_R\partial\beta_L-\alpha_L\partial\beta_R)\\
&\qquad+2m(\alpha_R\alpha_L+\beta_R\beta_L)-2t_2\alpha_L\beta_R,
\end{split}\label{eq:H_con}
\end{equation}
As the name suggests, $\alpha$ ($\beta$) with subscript $R/L$ represent the right-/left- mover on the top (bottom) leg.
When $m=0$, only the fields $\alpha_L$ and $\beta_R$ are coupled (by the $t_2$ term) and thus get gapped out, and we are left with one pair of chiral gapless modes, namely $\alpha_R$ on the top leg and $\beta_L$ on the bottom leg. For $m\neq 0$, all the modes are gapped out and there is no edge mode.

To be able to characterize the spatial structure of the chiral modes using numerical methods we consider the expectation of the thermal current operators in the ground state. The chiral modes do not carry well defined charge but since the energy is conserved they do carry heat current. The latter can be evaluated in both non-interacting and fully interacting models using DMRG.
This approach will allow us to understand the interacting model where the microscopic derivation of the low energy Hamiltonian as in Eq.~(\ref{eq:H_con}) is not readily available. We obtain the thermal current operators on each leg by partitioning the system into two parts (L/R) and going to the decoupled and massless limit $t_{1,2}=0, m=0$\footnote{If we do not set $m=0$, there will be an extra term in the thermal currents, corresponding to the interacting terms in the later sections. With such terms, however, the currents change sign as $m$ increases, prompting the removal thereof.}
\begin{equation}
\begin{split}
&I_{Q,\alpha}=-i\langle[H_{\alpha,L},H]\rangle=2it^2\langle\alpha_{-2}\alpha_0\rangle,\\
&I_{Q,\beta}=-i\langle[H_{\beta,L},H]\rangle=2it^2\langle\beta_{-2}\beta_0\rangle.
\end{split}\label{eq:I}
\end{equation}
The relation between the currents and $m$ is shown in Fig.~\ref{fig:2}(a), where we take $t=t_1=2t_2=1$. We see that the thermal currents evaluated in the ground state clearly capture the separation of the left- and right-movers. In the non-interacting limit thermal current can be evaluated at finite temperature to show an additional universal contribution to the ground state current, $\langle I_{Q}(T)\rangle = I_{Q}(T=0) + \frac{c\pi}{12}k_{B}T^{2}$, where $c=1/2$ is the central charge for chiral Ising mode [\onlinecite{cappelli2002}].

The above construction is naturally generalized to multi-leg ladders. The Hamiltonian for a ladder with $N$ legs can be written as
\begin{equation}\label{ni1}
\begin{split}
H&=i\sum_{j,l}(-1)^{l+1}(t+m(-1)^j)\alpha_{j,l}\alpha_{j+1,l}\\
&\qquad+it_1\sum_{j,l}(-1)^l\alpha_{j,l}\alpha_{j,l+1}\\
&\qquad+it_2\sum_{j,l}(\alpha_{j,l}\alpha_{j+1,l+1}-\alpha_{j,l+1}\alpha_{j+1,l}),\\
\end{split}
\end{equation}
where $l$ extends from 1 to $N$.
To keep the notation compact we use $\alpha_{j,l}$ to denote the Majorana operator on the site $j$ of the $l$th chain. For periodic boundary conditions in both directions the model defined in Eq.~(\ref{ni1}) can be readily solved by transforming to the momentum space representation, see Appendix \ref{sec:appendix}. The resulting Bloch Hamiltonian is a $2\times 2$ hermitian matrix and has a low-energy spectrum of the form
\begin{equation}\label{ni2}
\epsilon_{\bq}=\pm4\sqrt{t^2q_x^2+t_1^2q_y^2+(m-2t_2\cos k_y)^2}.
\end{equation}
Here ${\bq}$ is the crystal momentum relative to the nodal points ${\bk}=(0,0),(0,\pi)$. The spectrum is fully gapped except when $2|t_2|=|m|$.  Two gapped phases occurring for  $2|t_2|>|m|$ are topological and have a single gapless chiral Majorana mode at the boundary with the chirality determined by $\sgn{t_2}$.  The other two gapped phases for $2|t_2|<|m|$ are topologically trivial. 

The $N$-leg ladder defined by Hamiltonian Eq.~(\ref{ni1}) provides an excellent primer for the coupled wire construction. As we already mentioned the single chain and two-leg ladder exhibit an isolated critical point at which a pair of gapless Majorana modes exist. Already for $N=2$ these become spatially separated on two legs. For $N>2$ this critical point begins to expand into a critical phase. This is illustrated  in Fig.~\ref{fig:2}(d) which shows the excitation energy for an $N$-leg ladder with open boundary conditions along the direction perpendicular to the legs (i.e.\ the periodic strip geometry) as a function of $m$. We observe that the gapless region broadens with the increasing $N$ and for $N=20$ it spans almost the entire width of the  topological phase defined by $|m|<2t_2$. In the topological phase gapless modes occur at the edge, whereas in the trivial phase there are no gapless edge modes and the spectrum is gapped.    

Thermal currents  on each leg for the $N=20$ case 
\begin{equation}
I_{Q,l}=2it^2\langle\alpha_{-2,l}\alpha_{0,l}\rangle.
\end{equation}
are shown in Fig.~\ref{fig:2}(c). They indicate a clear spatial separation of the chiral modes to the edges of the system with vanishingly small currents in the gapped bulk.

\section{The interacting model}

With the intuition gained from the non-interacting model, we now focus on the full interacting model, Eq.~(\ref{eq:H_ladder}). As the mean-field theory suggests, we expect a TCI phase transition to occur in the ladder just as it does in the chain model. Moreover, as the mean-field model has edge modes spatially separated on the two edges, we expect this to also hold in the full model. We will show numerical evidence for this mode separation  below. This result allows an interesting new way of viewing the interacting model: instead of a detour into the non-interacting models, we start from two copies of the GSV chain model tuned to the TCI point and seek to gap out the bulk modes by interchain coupling. On each isolated chain, there will be a pair of left- and right-moving gapless modes characterized by the TCI CFT with central charge $c=7/10$. As the two chains get coupled, one pair of oppositely-moving modes are gapped out, leaving us with another pair residing on the two edges.

\begin{figure}
    \centering
    \includegraphics[width=0.48\textwidth]{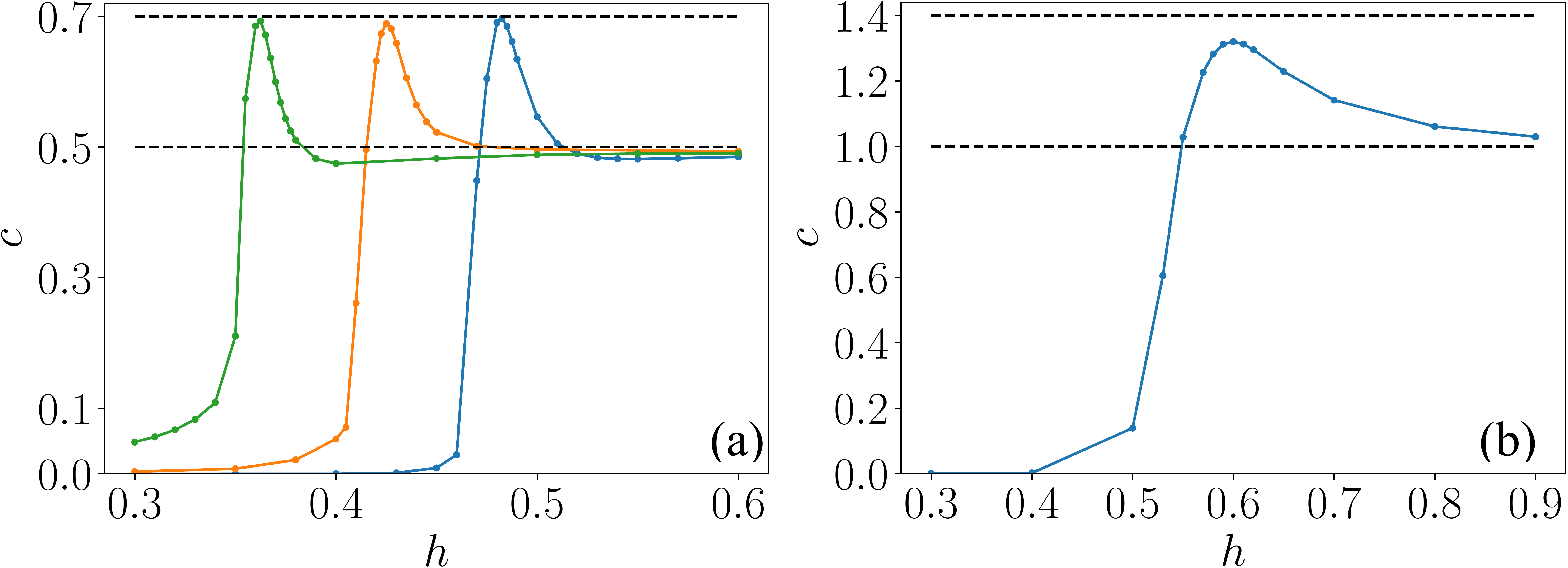}
    \caption{The central charge for the interacting two-leg ladder with (a) $t_1=2t_2=0.6\ (\mathrm{blue}),1\ (\mathrm{orange}),2\ (\mathrm{green})$ and (b) $t_1=t_2=0$. We fix $g=1$ and $J=0.3$. }
    \label{fig:central_charge}
\end{figure}

TCI phase transition is numerically confirmed by calculating the central charge $c$ using DMRG as a function of parameter $h$ which is known to drive the transition in the chain model. We apply the standard method for computing $c$ by  fitting the formula [\onlinecite{Calabrese2004}]
\begin{equation}
S_A=\frac{c}{3}\ln \Big(\frac{L}{\pi a}\sin\frac{\pi l_A}{L}\Big)+S_0,
\end{equation}
where $a$ is the lattice constant, $l_A$ and $L$ are the subsystem and total length, and $S_0$ is a non-universal constant. $S_A$ denotes the entanglement entropy of subsystem $A$, defined by
\begin{equation}
S_A=\mathrm{Tr}(\rho_A\ln\rho_A),
\end{equation}
where the reduced density matrix $\rho_A$ is given by $\rho_A=\mathrm{Tr}_B\rho$. The results are shown in Fig.~\ref{fig:central_charge}. We show the results for $t_1=2t_2=0.6,1,2$ in Fig.~\ref{fig:central_charge}(a) and for the decoupled case $t_1=t_2=0$ in Fig.~\ref{fig:central_charge}(b). In the latter case, the central charges from the two chains add up and give twice the value as in Ref.~\onlinecite{Grover2014}, as expected for two decoupled chains undergoing the TCI transition.  When the chains are coupled, however, the central charge behavior remains qualitatively the same as in the single chain model, indicating only one pair of the left- and right-moving chiral modes with $c=7/10$. It is worth noting  that the critical field $h_c$ decreases as $t_{1,2}$ increase.

\begin{figure}
    \centering
    \includegraphics[width=0.48\textwidth]{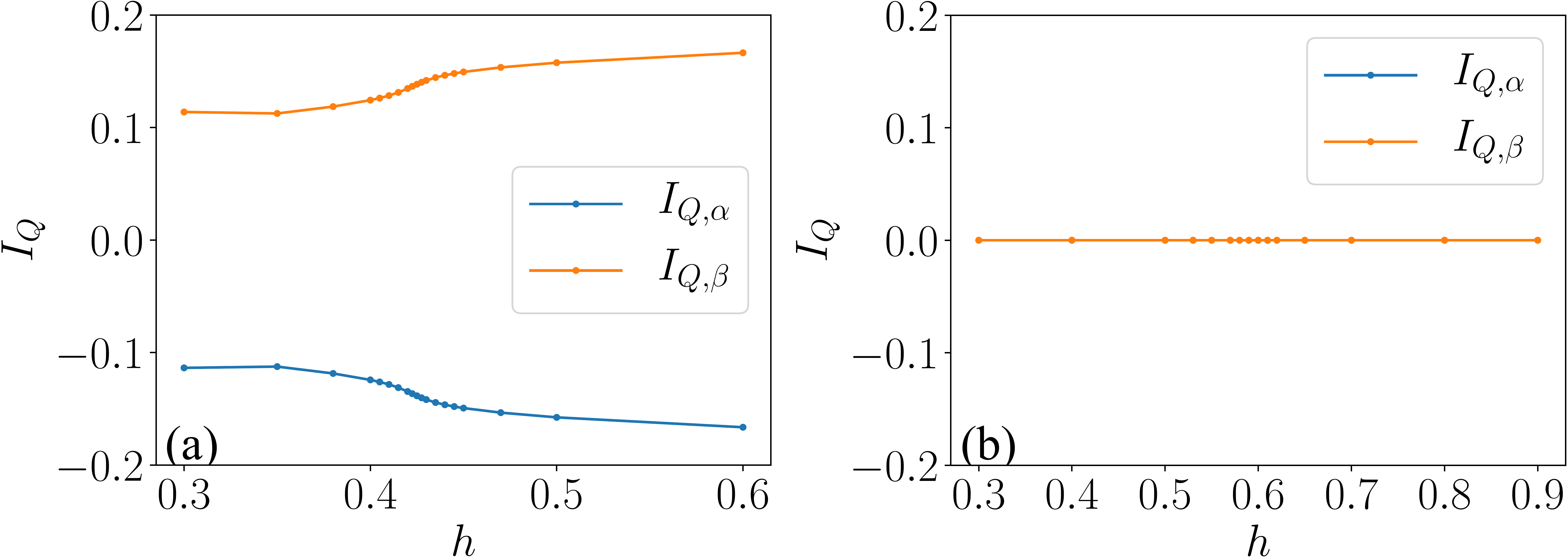}
    \caption{The thermal currents in the fully interacting two-leg ladder (computed using DMRG) for (a) $t_1=2t_2=1$ and (b) $t_1=t_2=0$. }
    \label{fig:thermal_current}
  \end{figure}

To identify the chirality and the spatial structure of the edge modes we calculate the chain-resolved thermal currents defined in Eq.~(\ref{eq:I}) using DMRG. The results are shown in Fig.~\ref{fig:thermal_current}
and support the conjecture that $c=7/10$ modes are spatially separated on the ladder and can plausibly be viewed as a precursor of the chiral edge modes in a 2D system.

The interacting Hamiltonian can be generalized to $N$ legs. In analogy with the non-interacting $N$-leg model discussed in Sec.\ II we expect the TCI gapless edge modes to be stabilized over a range of parameter $h$ that widens with  increasing $N$, c.f.\ Fig.~\ref{fig:2}(d). Similarly we expect the thermal currents associated with the $c=7/10$ gapless modes to be segregated to the edges and the corresponding 2D topological phase will be stable with increasing $N$ just like in the flux ladders [\onlinecite{Strinati2019}].

\begin{figure*}
    \centering
    \includegraphics[width=0.95\textwidth]{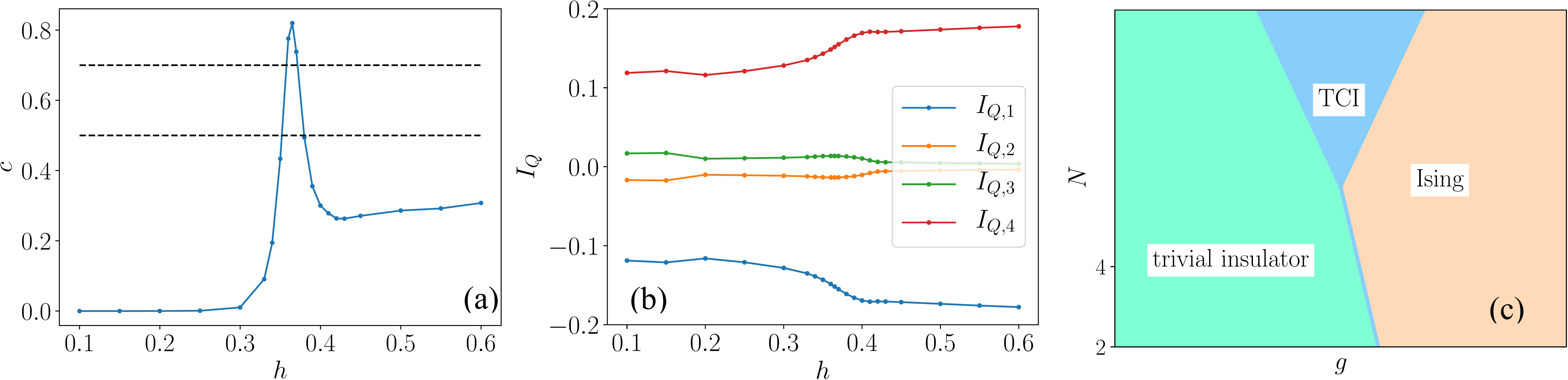}
    \caption{(a) The central charge and (b) the thermal currents in the interacting four-leg ladder for $t_1=2t_2=1$. (c) The tentative phase diagram in the $N$-$g$ space based on the numerical and analytical analyses. }
    \label{fig:4l_int}
\end{figure*}

In practice a DMRG simulation for the four-leg fully interacting case is already computationally very heavy, and we therefore restrict our numerical  calculations to this size.  The results for central charge $c$ and thermal currents are displayed in Fig.~\ref{fig:4l_int}(a, b). They show the same qualitative features as the two-leg ladder: a peak in the central charge close to $h_c=0.36$ suggests a phase transition and the distribution of thermal currents confirms that gapless modes are localized at the outer two legs of the 4-leg ladder which can be viewed as forming the edge of the system.
Quantitatively, however, the peak value of the central charge is closer to 0.8 than $7/10$ expected for the TCI transition. Also, for $h>h_c$, the central charge does not convincingly saturate at $1/2$ as one might expect in this regime. We therefore suspect that, despite our significant computational effort, the DMRG results for the 4-leg case have not fully converged to provide a quantitatively reliable result for the central charge. Based on the numerical and analytical results we have, we propose a tentative phase diagram as one goes to 2D in Fig.~\ref{fig:4l_int}(c).

\section{Possible realization in arrays of Majorana Cooper pair boxes}
We propose a geometry to realize the 2D TCI phase in networks of Majorana Cooper pair boxes (MCBs)~[\onlinecite{hassler2012strongly,barkeshli2015physical,plugge2016roadmap,karzig2017scalable}], superconducting islands harboring Majorana fermions. Such boxes are realized by introducing a superconducting island hosting a semiconducting nanowire. By tuning a magnetic field and chemical potential, there are Majorana fermions localized on both edges of the nanowire.  

\begin{figure}
\centering \includegraphics[width=0.5\columnwidth]{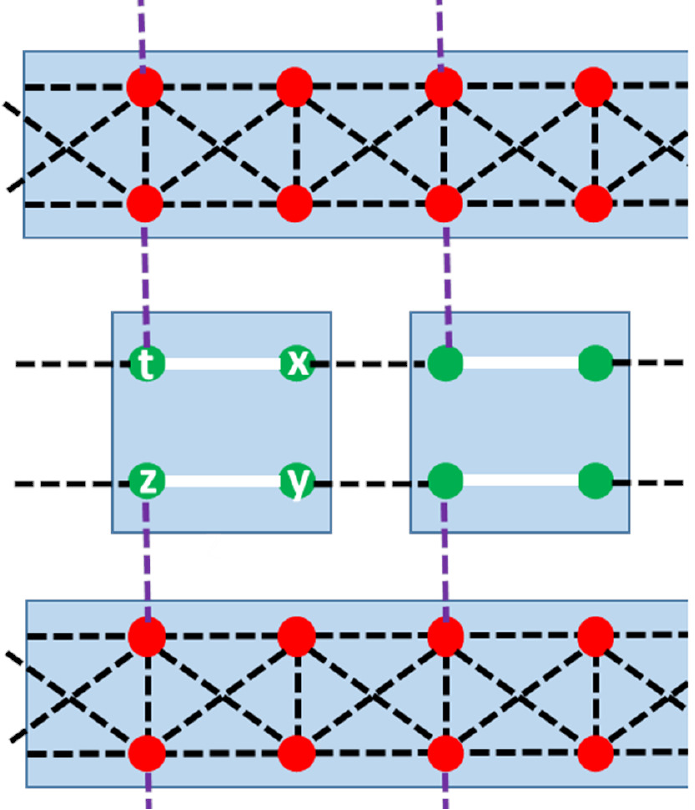}
\caption{
A geometry that realizes the 2D TCI topological phase, consisting of 1D arrays of two types of Majorana Cooper pair boxes (MCBs) in alternating pattern. The first type is a large superconductor which harbors Majorana fermions (red dots) by placing nanowires on the top (for the sake of simplicity we omit drawing the nanowires). By tuning couplings between the Majorana fermions, the island renders the $2$-leg Majorana ladder system discussed above. The second type  is the alignment of MCBs each of which has four Majorana fermions (green dots) by putting two nanowires (white lines) on the top. It reproduces the transverse field Ising model by tuning a strong charging energy and tunneling between the Majorana fermions. The indices, $t,x,y,z$ correspond to the superscript of the Majorana fermion operators defined above Eq.~(\ref{tu}). The dotted lines depict the coupling between the Majorana fermions.  } \label{mzm} 
\end{figure}

To generate the desired TCI phase, we consider a geometry where two types of 1D alignment of MCBs are placed in an alternating pattern as shown in Fig.~\ref{mzm}. The adjacent two 1D alignments are coupled via tunneling of Majorana fermions. For a moment, we turn off this tunneling and focus on the decoupled two types of the 1D alignment. \par
The first one is a large superconducting island with nanowires which gives rise to arrays of Majorana fermions (red dots in Fig.~\ref{mzm}). By tuning couplings between these Majorana fermions, we render the Hamiltonian of $2$-leg ladder, $H_M$ described in Eq.~(\ref{eq:H_ladder}).
The second alignment is a network of MCBs each of which has two nanowires, yielding four Majorana fermions. In the same spirit of Ref.~\onlinecite{glazman1997new}, where a spin chain is realized in arrays of Cooper pair boxes, we can construct the transverse field Ising model~[$H_s$ in Eq.~(\ref{eq:H_GSV})] in this geometry.
Labeling the four Majorana fermions by $\gamma^{k}_{i,j}$ $(k=t,x,y,z)$ (the subscript index denotes the 2D coordinates), the Hamiltonian of the MCB network reads
\begin{equation}
\begin{split}
H=&\sum_{i,j}-ih\gamma^t_{i,j}\gamma^{x}_{i,j}+it\gamma^x_{i,j}\gamma^t_{i+1,j}+it\gamma^y_{i,j}\gamma^z_{i+1,j}\\
&-U\gamma^t_{i,j}\gamma^x_{i,j}\gamma^y_{i,j}\gamma^z_{i,j}.\label{tu}
\end{split}
\end{equation}
The first three terms describe hopping between Majorana fermions and the last term comes from a charging energy on each island. When the charging energy is larger than other hopping terms, \emph{i.e.,} $U\gg t$, we obtain the following constraint on fermion parity: 
\begin{equation}
   \gamma^t_{i,j}\gamma^x_{i,j}\gamma^y_{i,j}\gamma^z_{i,j}=1,\label{hitagi}
\end{equation}
which is reminiscent of a gauge fixing condition discussed in Ref. \onlinecite{Kitaev2006}. This constraint allows us to define a spin-$1/2$ operator by
\begin{equation}
    \mu^k_{i,j}=i\gamma^{t}_{i,j}\gamma^{k}_{i,j}\;\;(k=x,y,z).\label{neko}
\end{equation}
Since $U\gg t$, we regard the second and third terms in Eq.~(\ref{tu}) representing hopping between Majorana fermions of adjacent islands as perturbation. The second order perturbation analysis shows that a such hopping between the adjacent islands has the form~$ J(\gamma^x_{i,j}\gamma^t_{i+1,j})(\gamma^y_{i,j}\gamma^z_{i+1,j})$ which is equivalent to $J\mu^z_{i,j}\mu^z_{i+1,j}$ by use of Eqs.~(\ref{hitagi})(\ref{neko}), where $J$ is proportional to $t^2/U$. Noting that the first term in Eq.~(\ref{tu}) is transformed to $-h\mu^x_{i,j}$, it follows that Hamiltonian~(\ref{tu}) reproduces the transverse field Ising model. \par
Now we turn on the tunneling between adjacent 1D MCBs in the perpendicular direction. Such a tunneling is described by (see also the purple dashed lines in Fig.~\ref{mzm})
\begin{equation}
  H_{\perp}=\sum_{i,j} it_{\perp}\beta_{2i+1,j-1}\gamma^t_{i,j}+it_{\perp}\alpha_{2i+1,j+1}\gamma^z_{i,j}.\nonumber \end{equation}
Assuming $U\gg t_{\perp}$, we can again resort to the second order perturbation analysis, leading to the term 
\begin{equation}
    \lambda \mu^z_{i,j}\beta_{i,j-1}\alpha_{i,j+1},\label{kambara}
\end{equation}
 where $\lambda\sim t^2_{\perp}/U$ and we have used Eq.~(\ref{neko}). When $t_1=2t_2$ in each $2$-leg Majorana ladder, the preceding discussion around Eqs.(\ref{H_ladder0})-(\ref{eq:H_con}) shows that there are a pair of right/left moving gapless modes in each ladder, furthermore, these modes are spatially decoupled; the wave function of the right/left moving mode is localized along the  top/bottom of the ladder. Hence, similarly to Eqs.~(\ref{5}) and (\ref{6}), the local Majorana fermion $\beta_{i,j-1}$ and $\alpha_{i,j+1}$ in Eq.~(\ref{kambara}) transmutes to a left and right moving gapless Majorana fields defined by $\beta_{L,j-1}$ and $\alpha_{R,j+1}$ in the continuum limit. We use a well-known fact that a continuum field theory description of the transverse field Ising model is given by a standard $\varphi^4$-theory~[\onlinecite{polyakov1989gauge}]. Indeed, if a mass term $r\varphi^2$ is added to this theory, the theory flows to gapped phase (spontaneously broken phase) when $r>0\;(r<0)$, which corresponds to disorder (order) phase of the transverse field Ising model. Based on this fact,  the potential of the effective field theory of the networks of the transverse field Ising model together with the tunneling term in Eq.~(\ref{kambara}) between adjacent $2$-leg Majorana ladder systems has the form
 \begin{equation}
     \sum_{j}i\beta_{L,j-1}\alpha_{R,j+1}\varphi_{j}+\varphi^4_{j}+r\varphi^2_{j}.\label{tsubasa} 
 \end{equation}
Therefore, when $t_1=2t_2$, effective field theory of each
MCBs of the transverse field Ising model with bottom/top chain of the $2$-leg ladder above/below the MCB manifests the one of the 1D GSV model~(\ref{eq:H_GSV}).  By tuning the coupling $h$ in Hamiltonian~(\ref{tu}), which is physically implemented via adjusting a chemical potential of a nanowire, correspondingly, tuning the mass term $r$ in Eq.~(\ref{tsubasa}), we obtain decoupled arrays of the $1+1$-dimensional TCI CFTs.\par
If we relax the condition of $t_1=2t_2$, the coupling between the gapless Majorana field is induced, \emph{i.e.,} the term $iu_{\perp}\beta_{L,j-1}\alpha_{R,j+1}$ is generated, where $u_{\perp}$ is proportional to $t_1-2t_2$. By virtue of the coupled wire construction, this situation closely parallels the $N$-leg ladder system that we believe constructs the desired 2D TCI topological phase as the adjacent TCI CFTs are coupled via Majorana couplings.

\section{Conclusions}

The coupled wire construction provides an intuitive approach to assembling strongly
interacting phases in two dimensions from well understood 1D components. In this work we made an attempt to construct in this way a stable 2D topological phase whose bulk is fully gapped and supports chiral gapless edge modes described by tricritical Ising conformal field theory with central charge $c=7/10$. This critical system is interesting because it exhibits supersymmetry and its primary fields include Fibonacci anyon which
is known to permit universal topological quantum computation.

Starting from the Grover-Sheng-Vishwanath chain model, which is known to support the TCI CFT at its critical point, we formed $N$-leg ladders and analyzed them using DMRG. The results for $N=2,4$  indicate robust separation of the gapless modes to the opposite edges of the ladder as measured by thermal current expectation values. Analysis of the central charge behavior based on the entanglement entropy formula shows convincing evidence for the TCI CFT in the 2-leg ladder. In the 4-leg ladder the behavior is qualitatively similar but quantitatively less convincing, presumably because  the large bond dimension prevents us from obtaining the fully converged DMRG results near the critical point.

In approaching the 2D limit it would be desirable to numerically study ladders with $N>4$. Unfortunately, with available computational resources we were not able to perform DMRG for such systems. Our results for $N=2,4$ are strongly suggestive if not entirely conclusive that the model for large $N$ describes a gapped 2D topological phase with protected gapless TCI boundary modes.
Another possible direction is to calculate the finite temperature thermal currents, which, unlike the ground state ones, are expected to be directly related to the central charge, and thus give yet another probe into the nature of the phases. We hope that our work here will motivate future studies of this intriguing problem.

\section{Acknowledgements}
We thank O{\u g}uzhan Can, {\' E}tienne Lantagne-Hurtubise, Miles Stoudenmire, Tarun Tummuru, Rohit R. Kalloor, and Alexei M. Tsvelik for helpful discussions. The work described in this article was supported by NSERC, CIfAR, and CFREF. C. L. was also supported by the QuEST scholarship at the University of British Columbia. Part of the results were included in the master's thesis of C. L. The DMRG calculations were performed using the ITensor library [\onlinecite{fishman2020itensor}]. This work was partially supported by the European Union’s Horizon 2020 research and innovation programme (Grant Agreement LEGOTOP No. 788715), the DFG (CRC/Transregio 183, EI 519/7-1),  Israel Science Foundation (ISF) and MAFAT Quantum Science and Technology grant, an NSF/DMR-BSF 2018643 grant, and Koshland postdoc fellowship.

\appendix
\section{Details on the non-interacting model}\label{sec:appendix}
\begin{figure}
    \centering
    \includegraphics[width=0.45\textwidth]{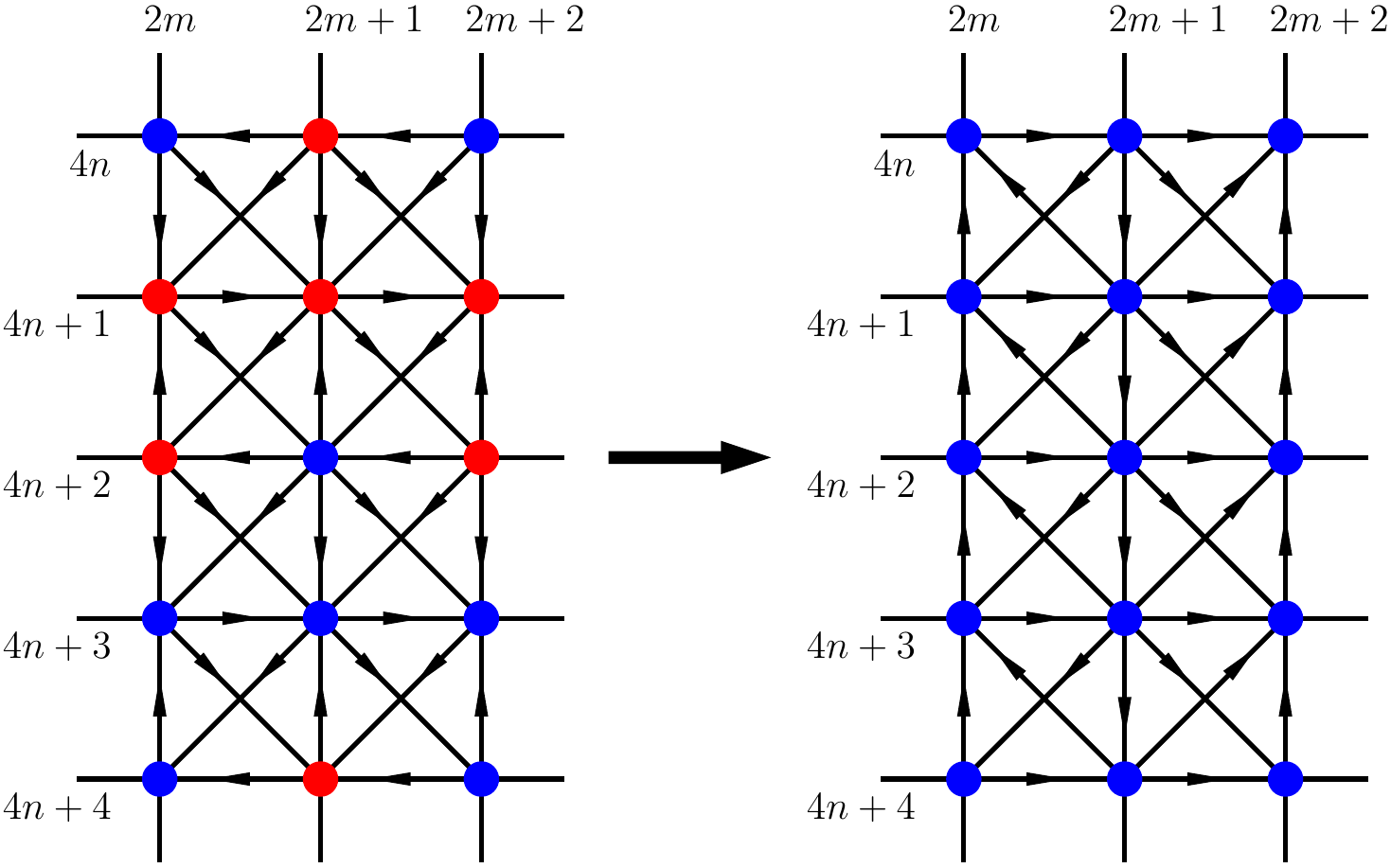}
    \caption{The gauge transformation in Eq.~(\ref{eq:gauge}). We introduce a minus sign for the red sites so that the unit cell comprises only two sites.}
    \label{fig:gauge}
\end{figure}

In this appendix we discuss in more detail the non-interacting model defined in Eq.~(\ref{ni1}). While at first sight the phase factors stipulate a $2\times 2$ unit cell and thus require solving a $4\times 4$ matrix in the momentum basis, we can exploit the gauge degree of freedom, where the Majorana operators are defined up to a sign, and set $\alpha_{j,l}\rightarrow-\alpha_{j,l}$ for the following sites:
\begin{equation}
\begin{split}
&j=2m, l=4n+1;\\
&j=2m, l=4n+2;\\
&j=2m+1, l=4n;\\
&j=2m+1, l=4n+1;
\end{split}\label{eq:gauge}
\end{equation}
see Fig.~\ref{fig:gauge}. The Hamiltonian will then be
\begin{equation}
\begin{split}
H&=i\sum_{j,l}(t+m(-1)^j)\alpha_{j,l}\alpha_{j+1,l}+t_1(-1)^{j+1}\alpha_{j,l}\alpha_{j,l+1}\\
&+it_2\sum_{j,l}(-1)^{j+1}(\alpha_{j,l}\alpha_{j+1,l+1}+\alpha_{j,l+1}\alpha_{j+1,l})\\
&=i\sum_{j,l}\big((t+m)\alpha_{j,l}\beta_{j,l}+(t-m)\beta_{j,l}\alpha_{j+1,l}\big)\\
&+it_1\sum_{j,l}(-\alpha_{j,l}\alpha_{j,l+1}+\beta_{j,l}\beta_{j,l+1})\\
&+it_2\sum_{j,l}(-\alpha_{j,l}\beta_{j,l+1}+\beta_{j,l}\alpha_{j+1,l+1}\\
&-\alpha_{j,l+1}\beta_{j,l}+\beta_{j,l+1}\alpha_{j+1,l}),
\end{split}
\end{equation}
where we defined
\begin{equation}
\alpha_{j,l}=\alpha_{2j,l},\ \beta_{j,l}=\alpha_{2j+1,l}.
\end{equation}

We introduce momentum-space Majorana operators
\begin{equation}
\begin{split}
\alpha_{j,l}=\sqrt{\frac{2}{N}}\sum_{j,l}\re^{i\br_{j,l}\cdot\bk}\alpha_{\bk},\\
\beta_{j,l}=\sqrt{\frac{2}{N}}\sum_{j,l}\re^{i\br_{j,l}\cdot\bk}\beta_{\bk},
\end{split}
\end{equation}
where $\br_{j,l}$ labels of the ($2\times1$) unit cell and the prefactors guarantee that $\alpha_\bk,\beta_\bk$ satisfies the usual anti-commutation relation. The hermiticity of the original Majorana operators translates to
\begin{equation}
\alpha_{-\bk}=\alpha_{\bk}^\dg,\ \beta_{-\bk}=\beta_{\bk}^\dg,
\end{equation}
and we only need to focus on half of the Brillouin zone (hBZ) due to particle-hole redundancy  [\onlinecite{Li2018}, \onlinecite{tummuru2020}]. The momentum-space Hamiltonian reads
\begin{equation}
H=\sum_{\bk\in\mathrm{hBZ}}\Psi_\bk^\dg H_\bk \Psi_\bk+E_0,
\end{equation}
with
\begin{equation}
\Psi_\bk=(\alpha_\bk,\beta_\bk)^T,
\end{equation}
\begin{equation}
H_\bk=2
\begin{pmatrix}
D_1 & D_2\\
D_2^* & -D_1\\
\end{pmatrix},
\end{equation}
and
\begin{equation}
\begin{split}
&D_1=2t_1\sin k_y,\\
&D_2=i(t+m)-i(t-m)\re^{-2ik_x}\\
&-it_2(\re^{ik_y}+\re^{-i(2k_x+k_y)}+\re^{-ik_y}+\re^{i(-2k_x+k_y)}),
\end{split}
\end{equation}
where $E_0$ is a constant which does not concern us here. We also set the original lattice constant to 1. As a Majorana model the Hamiltonian is particle-hole symmetric by construction, and the time-reversal symmetry is broken by the $t_2$ terms [\onlinecite{Rahmani2019}]. The Hamiltonian thus falls into class D in the ten-fold classification [\onlinecite{Chiu2016}]. The energy spectrum is given by
\begin{equation}
\begin{split}
E_{\bk,\pm}&=\pm2(D_1^2+|D_2|^2)^{1/2}\\
&=\pm4\big(t_1^2\sin^2 k_y+t^2\sin^2k_x\\
&+(m-2t_2\cos k_y)^2\cos^2 k_x\big)^{1/2}.
\end{split}
\end{equation}
We are interested in the physics around the Dirac point, where $E_{\bk,\pm}=0$. Solving for $\bk$ gives the condition 
\begin{equation}
k_x=0;k_y=0, m=2t_2\ \mathrm{or}\ k_y=\pi,m=-2t_2.
\end{equation}
Near the Dirac points $\bk_0 \in\{(0,0),(0,\pi)\}$, we expand the momentum as $\bk=\bk_0+\bq$, and the low-energy Hamiltonian is
\begin{equation}
\begin{split}
&H_{(0,0)}=4t_1q_y\sigma_z+4(2t_2-m)\sigma_y-4(2t_2+t-m)q_x\sigma_x,\\
&H_{(0,\pi)}=-4t_1q_y\sigma_z-4(2t_2+m)\sigma_y+4(2t_2-t+m)q_x\sigma_x,
\end{split}
\end{equation}
For $|m|\neq 2t_2$, rewriting $H_{(q_{x},q_{y})} = d_xq_x\sigma_x+d_yq_y\sigma_y+M\sigma_z$ at the Dirac points using an appropriate unitary transformation, we can calculate the Chern number from the following equation for this special case
\begin{align}
C=\frac{1}{2}\sum_{\mathbf{k}_0}\mathrm{sign}(d_xd_yM).
\end{align}
It follows that the Chern number is $-1$ for $|m|<2t_2$ and $0$ for $|m|>2t_2$ [\onlinecite{Bernevig}], consistent with Fig.~\ref{fig:2}(d).
\bibliography{biblio}
\end{document}